\documentclass[12pt,preprint]{aastex}
\usepackage{natbib}
\usepackage{graphicx}
\usepackage{amsmath}
\usepackage{epsfig}
\usepackage{graphicx}
\usepackage{rotating}
\usepackage{lscape}
\newcommand{\asec}{$^{\prime\prime}$}

\def\CNI{$^{13}$CN}
\def\CNII{C$^{15}$N}

\def\H{N$_{2}$H$^{+}$}
\def\D{N$_{2}$D$^{+}$}

\def\15N{$^{15}$NNH$^{+}$}
\def\N15{N$^{15}$NH$^{+}$}

\def\AMM{NH$_3$}

\def\METH{CH$_3$OH}

\def\HII{H{\sc ii}}

\def\kms{\mbox{km~s$^{-1}$}}

\def\cmq{cm$^{-2}$}

\def\Tex{\mbox{$T_{\rm ex}$}}

\begin{document}
\title{First measurements of $^{15}$N fractionation in N$_2$H$^+$ toward high-mass star forming cores
\thanks{Based on observations carried out with the IRAM-30m Telescope. IRAM is supported by INSU/CNRS (France), MPG (Germany) and IGN (Spain).}}
\author{F. Fontani$^{1}$, P. Caselli$^{2}$, A. Palau$^{3}$,  L. Bizzocchi$^{2}$, C. Ceccarelli$^{4,5}$}
\altaffiltext{1}{INAF-Osservatorio Astrofisico di Arcetri, L.go E. Fermi 5, Firenze, I-50125, Italy}
\altaffiltext{2}{Max Planck Institute for Extraterrestrial Physics, Giessenbachstrasse 1, D-85748 Garching, Germany }
\altaffiltext{3}{Centro de Radioastronom\'ia y Astrof\'isica, Universidad Nacional Aut\'onoma de M\'exico, PO Box 3-72, 58090 Morelia, Michoac\'an, M\'exico}
\altaffiltext{4}{Univ. Grenoble Alpes, IPAG, F-38000 Grenoble, France}
\altaffiltext{5}{CNRS, IPAG, F-38000 Grenoble, France}
%

\date{Received - ; accepted -}


%
%

\begin{abstract}
We report on the first measurements of the isotopic ratio $^{14}$N/$^{15}$N in 
N$_2$H$^+$ toward a statistically significant sample of high-mass star forming cores. 
The sources belong to the three main evolutionary categories of the high-mass star formation 
process: high-mass starless cores, high-mass protostellar objects, and 
ultracompact \HII\ regions. Simultaneous measurements of the 
$^{14}$N/$^{15}$N ratio in CN have been made. The $^{14}$N/$^{15}$N 
ratios derived from \H\ shows a large spread (from $\sim 180$ up to $\sim 1300$), 
while those derived from CN are in between the value measured in the
terrestrial atmosphere ($\sim 270$) and that of the proto-Solar nebula
($\sim 440$) for the large majority of the sources within the errors. 
However, this different spread might be due to the fact that
the sources detected in the \H\ isotopologues are more than those
detected in the CN ones. The $^{14}$N/$^{15}$N ratio does not change
significantly with the source evolutionary stage, which indicates that time seems
to be irrelevant for the fractionation of nitrogen.
We also find a possible anticorrelation between the $^{14}$N/$^{15}$N (as derived from \H ) 
and the H/D isotopic ratios. This suggests that $^{15}$N enrichment could
not be linked to the parameters that cause D enrichment, in agreement with the 
prediction by recent chemical models. These models, however, are not able 
to reproduce the observed large spread in $^{14}$N/$^{15}$N, pointing out
that some important routes of nitrogen fractionation could be still missing
in the models.
\end{abstract}


\section{Introduction}
\label{intro}

Nitrogen is the fifth most abundant chemical element in the universe, and 
exists in the form of two stable isotopes: $^{14}$N and $^{15}$N. 
In the terrestrial atmosphere (TA), the typical atomic composition 
$^{14}$N/$^{15}$N as derived from N$_2$ is $\sim 270$.
This value is quite consistent (within a factor 2) with that measured in 
cometary nitrile-bearing molecules (e.g.~Bockel\'ee-Morvan et al.~\citeyear{bockelee}, 
Manfroid et al.~\citeyear{manfroid}) 
and in primitive chondrites (e.g.~Briani et al.~\citeyear{briani}, 
Bonal et al.~\citeyear{bonal}), but it is significantly smaller than the value 
measured in the present-day Solar wind, which is $\sim 441 \pm 6$, as 
derived from the particles collected by the {\it Genesis} spacecraft 
(Marty et al.~\citeyear{marty}).
This value is comparable to that of the Jupiter's atmosphere 
(Fouchet et al.~\citeyear{fouchet04}), 
and it is considered to be representative of the so-called
"proto-Solar nebula" (PSN), from which our Sun has formed. 
However, to understand how these values are linked to the initial
chemical composition of the molecular environment in which
the System itself was formed, and put constraints on chemical models,
measurements of the $^{14}$N/$^{15}$N ratio in dense molecular star-forming
cores similar to that in which the Sun was likely formed are mandatory.

In dense molecular cores, where the direct observation of N$_2$ is
impossible, the $^{14}$N/$^{15}$N isotopic ratio is measured from abundant
N-bearing species such as \H , \AMM , CN, HCN, and HNC, and the results
are still puzzling. In fact, while the $^{14}$N/$^{15}$N ratio measured toward 
pre--stellar cores or protostellar envelopes in hydrogenated nitrogen compounds 
(such as \H\ and \AMM ) is comparable to the value found in the
PSN (e.g. in \AMM; Lis et al.~\citeyear{lis10}; Gerin et al.~\citeyear{gerin}; 
Daniel et al.~\citeyear{daniel13}), or even larger (1000$\pm$200 in N$_2$H$^+$; 
Bizzocchi et al.~\citeyear{bizzocchi13}), the $^{14}$N/$^{15}$N ratio derived 
from the nitrile-bearing species HCN and HNC is significantly lower (140 -- 360, 
Hily-Blant et al.~\citeyear{hilyblant13}; 120--400, Adande \& Ziurys~\citeyear{adande12}; 
160--290, Wampfler et al.~\citeyear{wampfler}). Differences in nitrogen hydrides 
and nitrile-bearing molecules within clouds are predicted by chemical models 
inclusive of spin-state chemistry (Wirstr\"{o}m et al.~\citeyear{wirstrom12}; 
Hily-Blant et al.~\citeyear{hilyblant13}), but no models have been able so far to 
reproduce the low $^{15}$N fractions measured in N$_2$H$^+$. 
The $^{15}$N fraction measured in CN toward starless cores has been 
found marginally consistent with the proto-Solar value of 441 
(500$\pm$75, Hily-Blant et al.~\citeyear{hilyblant13b}), in contrast with 
the large $^{15}$N fractions measured in HCN in both starless cores
(Hily-Blant et al.~\citeyear{hilyblant13b}) and protostellar objects
(Wampfler et al.~\citeyear{wampfler}). This is also expected 
from theory (Hily-Blant et al.~\citeyear{hilyblant13b}). 
A gradient of $^{14}$N/$^{15}$N with the Galactic distance has been measured by 
\citet{adande12} via observations of molecular clouds in CN and HCN, and 
found to be in agreement with predictions of Galactic chemical 
evolution models (e.g. Clayton~\citeyear{clayton03}, 
Romano \& Matteucci~\citeyear{rem03}).
However, the $^{15}$N fractionation of CN and HCN are often 
based on observations of the optically thin isotopologues containing 
$^{13}$C, because the main isotopologues are typically
optically thick, which makes it less accurate the derivation of the
column densities (e.g.~Hily-Blant~\citeyear{hilyblant13},~\citeyear{hilyblant13b}). 
Roueff et at.~(\citeyear{roueff15}) have found possible
reduced abundances of the $^{13}$C due to the fact that 
nitriles and isonitriles are predicted to be significantly depleted in 
$^{13}$C (Roueff et at.~\citeyear{roueff15}).
However, this depletion is estimated to be of the order of the $\sim 20\%$,
namely often of the order of the calibration uncertainties associated to
the measured column densities.

In star forming regions, D- and $^{15}$N- fractions are not correlated. 
For example, in L1544, N$_2$H$^+$ is highly deuterated (D-fraction $\sim$ 30\%; 
Caselli et al. 2002), while an anti-fractionation is observed in $^{15}$N 
(Bizzocchi et al.~\citeyear{bizzocchi13}). Wirstr\"{o}m et al.~(\citeyear{wirstrom12}) 
reproduced these differentiations and noticed that they are also present in meteoritic 
material, as only in a few cases $^{14}$N/$^{15}$N correlates with H/D ratios 
(Mandt et al.~\citeyear{mandt14}). This suggests an interstellar heritage of the 
Solar System $^{15}$N and D isotopic anomalies. Therefore, it is important to 
gather more data and to put stringent constraints on current chemical models, 
with the ultimate goal of investigating the possible link between cometary and interstellar 
medium material (see also Caselli \& Ceccarelli~\citeyear{cec12}), especially in intermediate- 
and high-mass dense cores, since growing evidence is suggesting that our 
Sun was born in a rich stellar cluster (Adams~\citeyear{adams10}). 

In this letter we report the first observational measurements of
the $^{14}$N/$^{15}$N ratio in a sample of dense cores associated with 
different stages of the high-mass star formation process, and already studied 
in deuterated molecules. Our study can investigate if the nitrogen fractionation 
process varies with the physical evolution of the core. What is more, the 
$^{14}$N/$^{15}$N ratio is derived from two well-studied nitrogen-bearing 
species, \H\ and CN, allowing us to investigate whether the fractionation of the 
two species follows different pathways as suggested by Hily-Blant 
et al.~(\citeyear{hilyblant13}).

\section{Observations}
\label{obs}

The spectra presented in this work are part of a spectral survey
performed with the IRAM-30m Telescope in two millimeter bands,
one centred at about 3~mm covering frequencies in the range 
89.11 -- 96.89 GHz, and another centred at about 1.3~mm covering 
frequencies in the range 216.0 -- 223.78 GHz, towards 26 dense cores 
divided into three main evolutionary categories: High-mass starless cores 
(HMSCs); high-mass protostellar objects (HMPOs); Ultracompact \HII\ regions 
(UC \HII s , see also Fontani et al.~\citeyear{fontani11} for details on the source 
selection criteria). In Fontani et al.~(\citeyear{fontani15}) we have published
part of these observations, namely that about the lines of \METH\ and of its 
deuterated forms, so that we refer to Sect.~2 and Table~2 of that work
for a description of the observational parameters and technical details. 
In the 3~mm band, the fundamental rotational transition of \H , \15N\ and
\N15\ (at 93173.4~MHz, 90263.8~MHz and 91205.7~MHz, respectively)
were observed simultaneously to the \METH\ observations described 
by Fontani et al.~(\citeyear{fontani15}) with velocity resolution of $\sim 0.65$ \kms\ 
(the telescope HPBW is $\sim 26$\asec ). In the 1~mm band, several hyperfine
components of the transition $N=2-1$ of radicals \CNI\ (50 components with
rest frequencies in between 217.0326 and 218.0311 GHz)
and \CNII\ (10 components with
rest frequencies in between 219.38749 and 219.93482 GHz) were
also observed with velocity resolution of $\sim 0.26$ \kms\
(the telescope HPBW is $\sim 11$\asec ).

\section{Results}
\label{res}

\subsection{\15N , \N15 , and \H }
\label{res_n2hp}

The \15N\ (1--0) has been detected in 14 cores: four HMSCs, six HMPOs and four
UC HIIs. 
Because the separation between the hyperfine components is comparable to the line widths, 
the hyperfine structure is unresolved in almost all spectra. Therefore, the lines have been
fit with single Gaussians, except in four cases (05358--mm3, 05358--mm1, ON1, 19410)
in which the fit to the hyperfine structure has given reliable results despite the
small separation of the components and the poor spectral resolution. 
This simplified approach could affect the determination of the line width, but
will not affect the measurement of the column densities because they will 
be derived from the total integrated area, as we will describe below.
The \N15\ (1--0) has been detected in 11 cores: two HMSCs, five HMPOs and 
four UC HIIs. The hyperfine components in this case are only three, with a faint 
central one and two fainter satellites, but the faintness of the satellites has 
forced us to fit many of the lines with Gaussians in this case too.
Finally, the \H (1--0) line is clearly detected in all sources and the fit to the hyperfine
structure gives good results in all spectra, pointing out that no hyperfine
anomalies are present.
In Fig.~\ref{spectra} we show the spectra of the \15N , \N15\ and \H\ (1--0)
lines in the representative core 05358--mm3.

The total column density of \H , averaged within the beam, has been
evaluated from Eq.~(A1) in Caselli et al.~(\citeyear{caselli02}). The formula
assumes the same \Tex\ for all the hyperfine components.  
For sources with total opacity well-constrained, \Tex\ was computed from
the output parameters obtained from the hyperfine fit procedure
(in particular the total optical depth, $\tau_{\rm t}$, and $T_{\rm ant}\times \tau_{\rm t}$)
following the approach described in the CLASS user 
manual\footnote{for details: http://iram.fr/IRAMFR/GILDAS/doc/html/class-html/class.html/}.
For the others, we have assumed the average \Tex\ derived from the
lines with well-constrained opacity in each evolutionary stage
(18, 43 and 57~K for HMSCs, HMPOs, and UC \HII s, respectively), 
and obtained the column density from Eq.~(A4) of Caselli et 
al.~(\citeyear{caselli02}), valid for optically thin lines. 
The spectral parameters used in our calculations ($S\mu^2$, Einstein coefficients,
rotational constants, energies and statistical weight of the levels) were taken
from the Cologne Database for Molecular Spectroscopy (M\"{u}ller et al.~\citeyear{muller}).

The total column densities of both \15N\ and \N15\ have been computed in
all detected sources from the total line integrated intensity with Eq.~(A4) of 
Caselli et al.~(\citeyear{caselli02}), assuming optically thin lines and
LTE conditions. As excitation temperature, we have used that derived 
from the \H (1--0) line. 
All column densities and the parameters used to derive them
(line integrated intensities and \Tex ), are given in Table~\ref{tab_res}.

\subsection{$^{13}$CN and C$^{15}$N}
\label{res_cn}

\begin{figure*}
\centerline{\includegraphics[angle=-90,width=18cm]{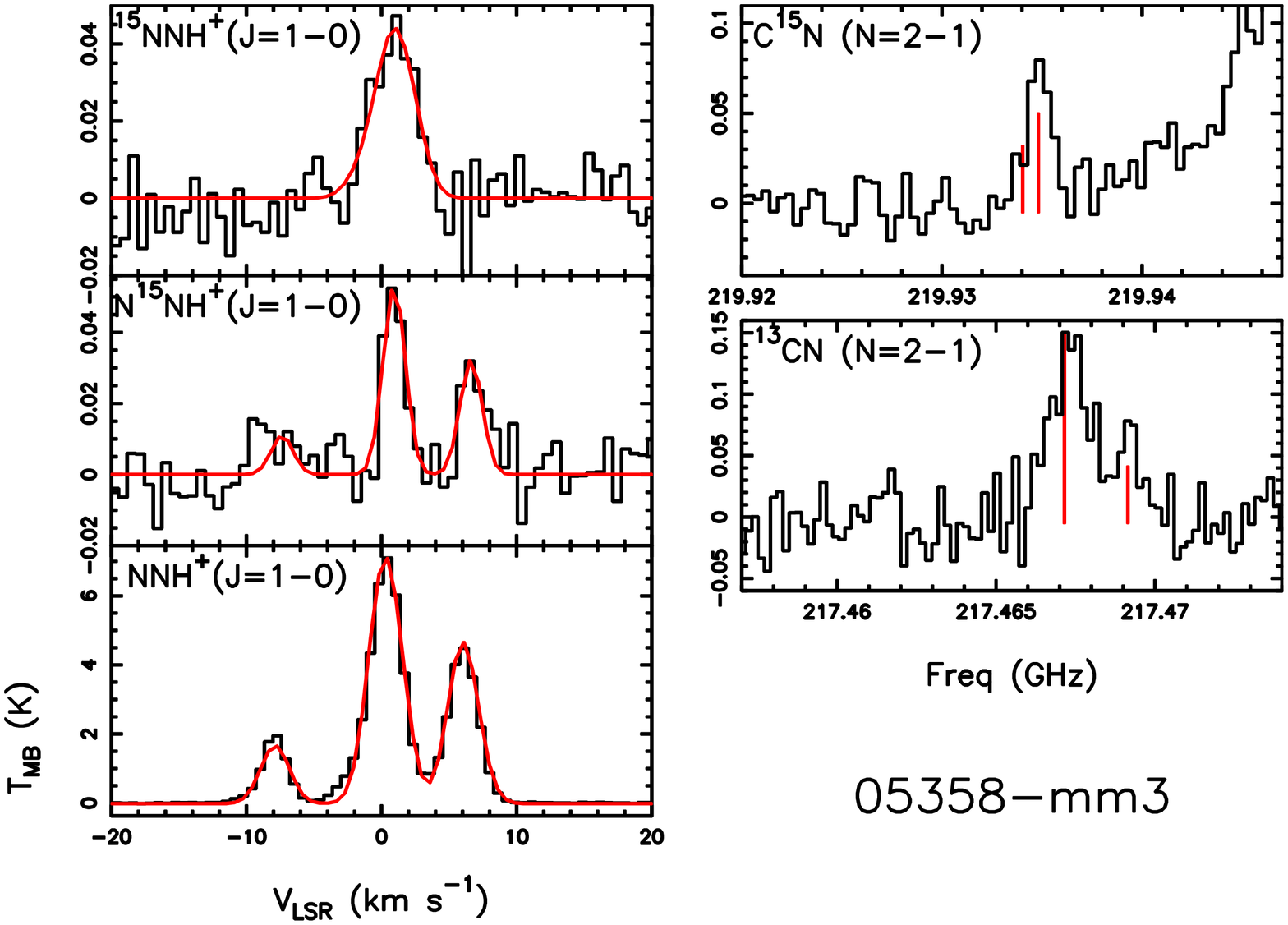}}
\caption[]{Left panels: spectra, from top to bottom, of \15N\ (1--0), \N15\ (1--0) and \H\ (1--0) 
towards 05358--mm3. The velocity interval shown is $\pm 20$ \kms\ from the
systemic velocity. For the \15N\ line, the red curve represents the best Gaussian fit,
while for the other lines corresponds to the best fit to the
hyperfine structure. 
\newline
Right panels: spectra of \CNII\ ($N=2-1$, top) and \CNI\ ($N=2-1$, bottom)
towards 05358--mm3 centred on the hyperfine components 
{mentioned in Sect.~\ref{res_cn}}, and indicated by the 
red vertical lines, the length of which is proportional to their relative
intensities. 
}
\label{spectra}
\end{figure*}

Several hyperfine components of the $N=2-1$ transitions of both \CNI\ and
\CNII\ were detected towards 13 cores, which yields a detection rate of 50$\%$. 
As expected, most of the cores were detected in the hyperfine components with 
larger values of $S\mu^{2}$, which are: 
$J,F_1,F=(5/2,3,4)-(3/2,2,3)$; $(5/2,3,3)-(3/2,2,2)$; $(5/2,3,2)-(3/2,2,1)$
for \CNI\ , and $J,F=(5/2,3)-(3/2,2)$; $(5/2,2)-(3/2,1)$ for \CNII . 
Representative spectra obtained towards core 05358--mm3 centred on these 
hyperfine components are shown in Fig.~\ref{spectra}.

To compute the column densities of both radicals, we have derived first the 
integrated intensity of the group of hyperfine components (Table~\ref{tab_res2})
mentioned above, and computed the total column densities from Eq.(A4) 
of Caselli et al.~(\citeyear{caselli02}), which assumes the same excitation 
temperature in the various rotational levels, and it is valid for optically thin lines.
The formula has been adapted to take the hyperfine structure into account, 
namely we have adopted the spectroscopic parameters (Einstein 
coefficients, energies and statistical weight of the levels)
of the hyperfine components considered only. 
All spectroscopic parameters, as well as the partition functions used to
derive the total column density, have been taken from the 
Cologne Database for Molecular Spectroscopy (M\"{u}ller et al.~\citeyear{muller}).
When the hyperfine components are not blended, their relative
intensities are consistent with the assumption of optically
thin lines and no hyperfine anomalies (within the calibration uncertainties).
As excitation temperature, we have used the \Tex\ reported in 
Table~\ref{tab_res}, derived from \H . This assumption
is critical, 
but the \CNII /\CNI\ column density ratio does 
not change by varying \Tex\ of even an order of magnitude, 
as it was already noted in other works dealing with isotopic ratios
(see e.g.~Sakai et al.~\citeyear{sakai12}, Fontani et al.~\citeyear{fontani14a}).

The CN column densities were then computed from the \CNI\ ones by applying the
isotopic ratio $^{12}$C/$^{13}$C derived from the relation between this ratio 
and the source galactocentric distance found for CN by Milam et al.~(\citeyear{milam05}).
Galactocentric distances have been taken from Fontani et al.~(\citeyear{fontani14b}).
We have not corrected the column densities for a filling factor because we do 
not have any direct estimate of the size of the emitting region of both molecules,
and we do not see any 
physical or chemical reason by which the emission of  \CNI\ and \CNII\ 
should arise from different regions.

In 19410+2336, and ON1, the \CNI\ hyperfine components mentioned
above are barely detected, while in G5.89--0.39 they are heavily affected by 
a nearby \METH\ stronger line. Therefore, in these sources we have used the 
better detected hyperfine component $J, F_1, F=(3/2,1,1)-(1/2 , 0, 1)$.
In G5.89--0.39 the blending problem affects also the \CNII\ lines, so that
we have used the isolated hyperfine component $J, F=(3/2,2)-(1/2,1)$ to
compute the \CNII\ column density. 
For I00117--MM1, I21307, I19035--VLA1, and 23033+5851, detected in \CNI\ 
but not in \CNII, we have derived an upper limit for the column density of
\CNII .
All column densities, and the line integrated intensities used to compute them,
are listed in Table~\ref{tab_res2}.

\begin{table*}
\normalsize
\begin{center}
\caption{Integrated intensity of the \15N, \N15\ and \H\ (1--0) transitions,
total beam-averaged column densities computed as explained in Sect.~\ref{res_n2hp},
and corresponding $^{14}$N/$^{15}$N isotopic ratios. Uncertainties in 
the isotope ratios have been computed from the propagation of errors 
on the column densities. These latters have been derived from the errors 
on the line areas for the optically thin lines, given by 
$rms \times \Delta v \times \sqrt{N}$ ($rms=$ root mean square noise in the spectrum,
$\Delta v =$ spectral velocity resolution, $N=$ number of channels with signal). For the
\H\ (1--0) lines with well-constrained opacity, we propagated the errors on 
$\tau_{\rm t}$ and $T_{\rm ant}\times \tau_{\rm t}$.
The source names are taken from Fontani et al.~(\citeyear{fontani11}).
}
\label{tab_res}
\tiny
\begin{tabular}{llllllllll}
\hline \hline
source  & \multicolumn{2}{c}{$^{15}$NNH$^+$(1--0)} &  \multicolumn{2}{c}{\N15 (1--0)} &  \multicolumn{2}{c}{\H (1--0)} & $\frac{{\rm N_2H^+}}{{\rm ^{15}NNH^+}}$ & $\frac{{\rm N_2H^+}}{{\rm N^{15}NH^+}}$ & \Tex  \\
        & $\int T_{\rm MB}{\rm d}v$ &  $N$(\15N ) & $\int T_{\rm MB}{\rm d}v$ & $N$(\N15 ) & $\int T_{\rm MB}{\rm d}v$ & $N$(\H ) & & & \\
	& K \kms & ($\times 10^{10}$\cmq ) & K \kms & ($\times 10^{10}$\cmq ) & K \kms & ($\times 10^{13}$\cmq ) & & & (K)\\
\cline{1-10}
\multicolumn{10}{c}{HMSC} \\
\cline{1-10}
 I00117--MM2        &  0.029(0.001)$^{t}$  & 5.68(0.09)   &  0.04        &  $\leq 6.8$  &  3.966(0.007)$^{a}$ & 3.8(0.5)   &670(98) & $\geq 561$        & 19\\
 AFGL5142--EC$^{w}$ &  0.040(0.002)   &   24(1) 	  &  0.04        &  $\leq 27$	&  8.81(0.01)$^{a}$  &  28(8)     &1100(360)& $\geq 1034$       & 76\\
 05358--mm3$^{w}$   &  0.17(0.01)     &   32(2) 	  & 0.20(0.01)   &    37(3)	& 39.69(0.05)  & 6.60(0.01) &210(12) & 180(13)               & 18\\  
  G034--G2          &  0.04           &   $\leq 6.95$	  & 0.04         &  $\leq 6.8$  & 4.314(0.001)$^{a}$ &  5.95(0.08)&$\geq 856$& $\geq 872$  &19 \\
  G034--F1          &  0.05           &  $\leq 6.94$	  &  0.04        &  $\leq 5.8$  & 4.4(0.1)$^{a}$     & 3.9(0.3)   &$\geq 566$& $\geq 672$  &11 \\
  G034--F2          &  0.04	      &   $\leq 7.83$	  & 0.04         &  $\leq 6.6$  & 9.20(0.03)   & 1.53(0.01) &$\geq 195$& $\geq 232$  &18 \\
  G028--C1          &  0.04	      &   $\leq 7.02$	  & 0.04         &  $\leq 5.9$  & 5.735(0.006)$^{a}$ & 8.5(0.1)   &$\geq 1217$&$\geq 1445$ &15 \\
 I20293--WC         &  0.063(0.006)   &   17(2) 	  &  0.08(0.01)  &    20(3)	&  7.150(0.003)$^{a}$ & 11.0(0.8) & 65(69)       & 550(98)        &29 \\
 22134--G$^{w}$     &  0.03	      &   $\leq 5.6$	  & 0.04         & $\leq 6.6$	& 7.80(0.03)   & 1.30(0.01) &$\geq 232$& $\geq 197$  & 18\\
 22134--B           &  0.04           &  $\leq 5.8$	  & 0.04         &   $\leq 5.7$ & 2.83(0.07)$^{a}$   & 1.8(0.3)   &$\geq 316$& $\geq 322$  & 14\\
\cline{1-10}      
\multicolumn{10}{c}{HMPO} \\
\cline{1-10}
  I00117--MM1   &  0.05(0.01)$^{t}$   &      14(3)	  & 0.05	 &   $\leq 12.8$&  2.88(0.01)$^{a}$  & 3.0(0.3)   & 220(60) &$\geq 233$       & 27  \\
  AFGL5142--MM  &  0.072(0.009)       &   32(4) 	  & 0.042(0.003)$^{t}$ &  18(1)	&  9.41(0.02)$^{a}$  &  23.5(1) & 740(97) &1300(210)             & 53  \\
 05358--mm1     &  0.17(0.02)         &   61(6) 	  & 0.18(0.02)   &  65(8)	& 36.63(0.05)  & 11.78(0.02)& 190(20) &180(23)              & 43  \\
 18089--1732    &  0.08(0.09)         &      22(9)	  & 0.10(0.01)   &   27(3)	&  9.312(0.002)$^{a}$ & 21(2)   & 1000(400) &800(100)	            & 31  \\ 
 18517+0437     &  0.058(0.008)       &    21(3)	  & 0.089(0.005) &   32(2)	& 25.98(0.03)  & 8.35(0.01) & 390(53) &260(20)              & 43  \\	  
   G75--core    &  0.06	              &     $\leq 22$	  & 0.05	 & $\leq 1.9$	& 13.95(0.06)  & 4.49(0.02) &$\geq 209$&$\geq 240$  & 43  \\
 I20293--MM1    &  0.086(0.007)       &   56(5) 	  & 0.19(0.02)   &  122(14)	& 16.410(0.001)$^{a}$ & 44.62(0.05)& 790(67) &370(42)              & 82  \\	 
  I21307        &  0.05	              &      $\leq 7.28$  & 0.05	 & $\leq 6.3$	&  2.13(0.05)$^{a}$  & 2.2(0.3)   & $\geq 301$ &$\geq 346$& 10  \\
  I23385        &  0.05	              &   $\leq 19$	  & 0.06	 & $\leq 23$	&  4.74(0.03)  & 1.53(0.01) & $\geq 81 $ &$\geq 66$ & 43  \\	 
  \cline{1-10}													      	  
\multicolumn{10}{c}{UC \HII } \\
\cline{1-10}
 G5.89--0.39   &  0.18(0.01)          &      84(6)	  & 0.13(0.01)   &  61(6)	& 66.3(0.5)    & 27.3(0.3)  & 320(25) &450(51)               & 57\\
 I19035--VLA1  &  0.06	              &      $\leq 31$    &  0.05	 & $\leq 24$	& 3.820(0.008)$^{a}$ & 18(1)      & $\geq 572$ &$\geq 754$ & 66\\	   
 19410+2336    &  0.18(0.02)          &       96(8)	  & 0.16(0.01)   &    82(8)	& 21.27(0.01)$^{a}$  & 43(5)   & 450(96) &500(100)               & 64\\
 ON1           &  0.21(0.01)          &     72(3)	  & 0.29(0.02)   &     94(8)	& 12.795(0.001)$^{a}$ &  33(4) & 460(22) &350(33)               & 39\\	
 I22134--VLA1  &  0.05	              &   $\leq 24$	  & 0.04	 &  $\leq 18$	&  4.71(0.03)  &  1.94(0.01)& $\geq 80$  &$\geq 108$ & 57\\
 23033+5951    &  0.048(0.007)        &     23(3)	  & 0.065(0.006) &   30(3)	&  7.660(0.001)$^{a}$ &  20(6)     & 900(400) & 700(250)               & 57\\	  
 NGC7538--IRS9 &  0.05	              &   $\leq 24$	  & 0.05	 & $\leq 21$	&  14.93(0.04) &  6.13(0.02)& $\geq 255$ &$\geq 294$ & 57\\
\hline
\end{tabular}
\end{center}
$^{a}$ lines with well-constrained optical depth, for which we do not list 
the integrated intensity but the parameter $T_{\rm ant}\times \tau_{\rm t}$, 
from which we have derived the column density as explained in Sect.~\ref{res_n2hp}; \\
$^{w}$ "warm" HMSC; \\
$^{t}$ tentative detection \\ 
\end{table*}
\normalsize

\begin{table*}
\normalsize
\begin{center}
\caption{Integrated intensity of the \CNII\ and \CNI\ transitions,
total beam-averaged column densities computed as explained in Sect.~\ref{res_cn},
and corresponding $^{14}$N/$^{15}$N isotopic ratio. 
We do not list the upper limits of the integrated intensity of \CNI\ and \CNII (2--1) 
for the sources undetected in both transitions. To compute $N$(CN), the \CNI\ 
column density was corrected by the $^{12}$C/$^{13}$C ratio at the source 
galactocentric distance, $D_{\rm GC}$ (Col.~5), assuming the relation between
$^{12}$C/$^{13}$C and $D_{\rm GC}$ derived for CN by Milam et al.~(\citeyear{milam05}). 
Uncertainties have been derived as explained in the caption of Table~\ref{tab_res}
for the optically thin lines.}
\label{tab_res2}
\scriptsize
\begin{tabular}{lllllll}
\hline \hline
source  & \multicolumn{2}{c}{C$^{15}$N(2--1)} &  \multicolumn{3}{c}{$^{13}$CN(2--1)} & $\frac{N({\rm CN})}{N({\rm C^{15}N})}$ \\
        & $\int T_{\rm MB}{\rm d}v$ & $N$(C$^{15}$N) & $\int T_{\rm MB}{\rm d}v$ & $D_{\rm GC}$ & $N$(CN) &  \\
	& K \kms & ($\times 10^{12}$\cmq ) & K \kms & kpc & ($\times 10^{14}$\cmq ) &  \\
\cline{1-7}
\multicolumn{7}{c}{HMSC} \\
\cline{1-7}
 I00117--MM2        &		  &	       &		  &	     &     &       \\
 AFGL5142--EC$^{w}$ & 0.14(0.02)$^{b}$ &   1.6(0.2) & 0.35(0.02)$^{a}$ & 10.3 & 5.3(0.3)  & 330(67)  \\
 05358--mm3$^{w}$   & 0.12(0.02)$^{b}$ &   0.7(0.1) & 0.36(0.02)$^{a}$ & 10.3 & 2.7(0.1)  & 400(90)  \\
  G034--G2          &		  &	       &		  &	      &    &       \\
  G034--F1          &		  &	       &		  &	      &     &      \\
  G034--F2          &		  &	       &		  &	      &    &       \\
  G028--C1          &		  &	       &	          &	      &     &      \\
 I20293--WC         &		  &	       &		  &	      &     &      \\
 22134--G$^{w}$     & 0.06(0.02)$^{b}$ &   0.3(0.1) & 0.16(0.02)$^{a}$ & 9.5 & 1.1(0.1)  & 330(150)  \\
 22134--B           &		  &	       &		  &	      &     &      \\
\cline{1-7}
\multicolumn{7}{c}{HMPO} \\
\cline{1-7}
  I00117--MM1   &$\leq 0.06$	     & $\leq 0.4$ & 0.13(0.02)$^{a}$ & 9.5 & 1.0(0.1)   &$\geq 250$ \\
  AFGL5142--MM  &0.28(0.02)$^{b}$ & 2.4(0.2)   & 0.40(0.02)$^{a}$ & 10.3 & 4.7(0.2)         & 190(23)  \\
 05358--mm1     &0.12(0.02)$^{b}$ & 1.5(0.2)   & 0.36(0.02)$^{a}$ & 10.3 & 3.7(0.2)	       & 240(40)  \\
 18089--1732    &0.14(0.03)$^{b}$ & 0.9(0.2)   & 0.80(0.02)$^{a}$ & 5.1 & 4.0(0.1)	       & 450(100)  \\
 18517+0437     &0.10(0.02)$^{b}$ & 0.8(0.2)  & 0.34(0.02)$^{a}$ & 6.5 & 2.4(0.2)	      & 310(80)   \\ 
   G75--core    &0.19(0.03)$^{b}$ & 1.4(0.2)   & 0.40(0.02)$^{a}$ & 8.4 & 3.4(0.2)          & 240(50)   \\
 I20293--MM1    &	      & 	   &		      & 		      &    &   \\     
  I21307        &$\leq 0.06$	     & $\leq 0.4$ & 0.09(0.02)$^{a}$ & 9.3 & 0.7(0.2)  &$\geq 175$  \\
  I23385        &	      & 	   &		      & 	              &    &   \\     
\cline{1-7}
\multicolumn{7}{c}{UC \HII } \\
\cline{1-7}
 G5.89--0.39   & 0.14(0.03)$^{d}$ &  3.5(0.8)  & 0.12(0.03)$^{c}$ & 7.2 & 12(3)	 & 350(160)      \\
 I19035--VLA1  & $\leq 0.06$	     & $\leq 0.6$ & 0.16(0.02)$^{a}$ & 7.0 & 1.6(0.2) &$\geq 270$ \\  
 19410+2336    & 0.07(0.02)$^{d}$ &  1.9(0.5)  & 0.07(0.02)$^{c}$ & 7.7 & 8(2)	 & 430(250)      \\
 ON1           & 0.07(0.02)$^{d}$ &  1.4(0.4)  & 0.06(0.02)$^{c}$ & 8.0   & 5(2)	 & 350(220)     \\        
 I22134--VLA1  & 0.09(0.02)$^{b}$ &  0.8(0.2)  & 0.18(0.02)$^{a}$ & 9.5 & 2.1(0.2)     & 250(84)      \\
 23033+5951    & $\leq 0.07$	     &  $\leq 0.7$& 0.21(0.02)$^{a}$ & 10.3 & 2.6(0.2)& $\geq 370$ \\
 NGC7538--IRS9 & 0.08(0.02)$^{b}$ &  0.7(0.2)  &0.14(0.02)$^{a}$  & 9.9 & 1.7(0.2)     & 230(90)      \\
\hline
\end{tabular}
\end{center}
$^{w}$ "warm" HMSC; \\
$^{a}$ from the hyperfine components $J, F_1, F =(5/2,3,4)-(3/2,2,3), (5/2,3,3)-(3/2, 2, 2)$
and $(5/2,3,2)-(3/2,2,1)$ ; \\
$^{b}$ from the hyperfine components $J,F=(5/2,3)-(3/2,2)$ and $J, F=(5/2,2)-(3/2,1)$; \\
$^{c}$ from the hyperfine component $J,F_1,F=(3/2,1,1)-(1/2, 0,1)$; \\
$^{d}$ from the hyperfine component $J,F=(3/2,2)-(1/2,1)$. \\
\end{table*}
\normalsize

\section{Discussion}
\label{discu}

\begin{figure*}
\centerline{\includegraphics[angle=-90,width=8cm]{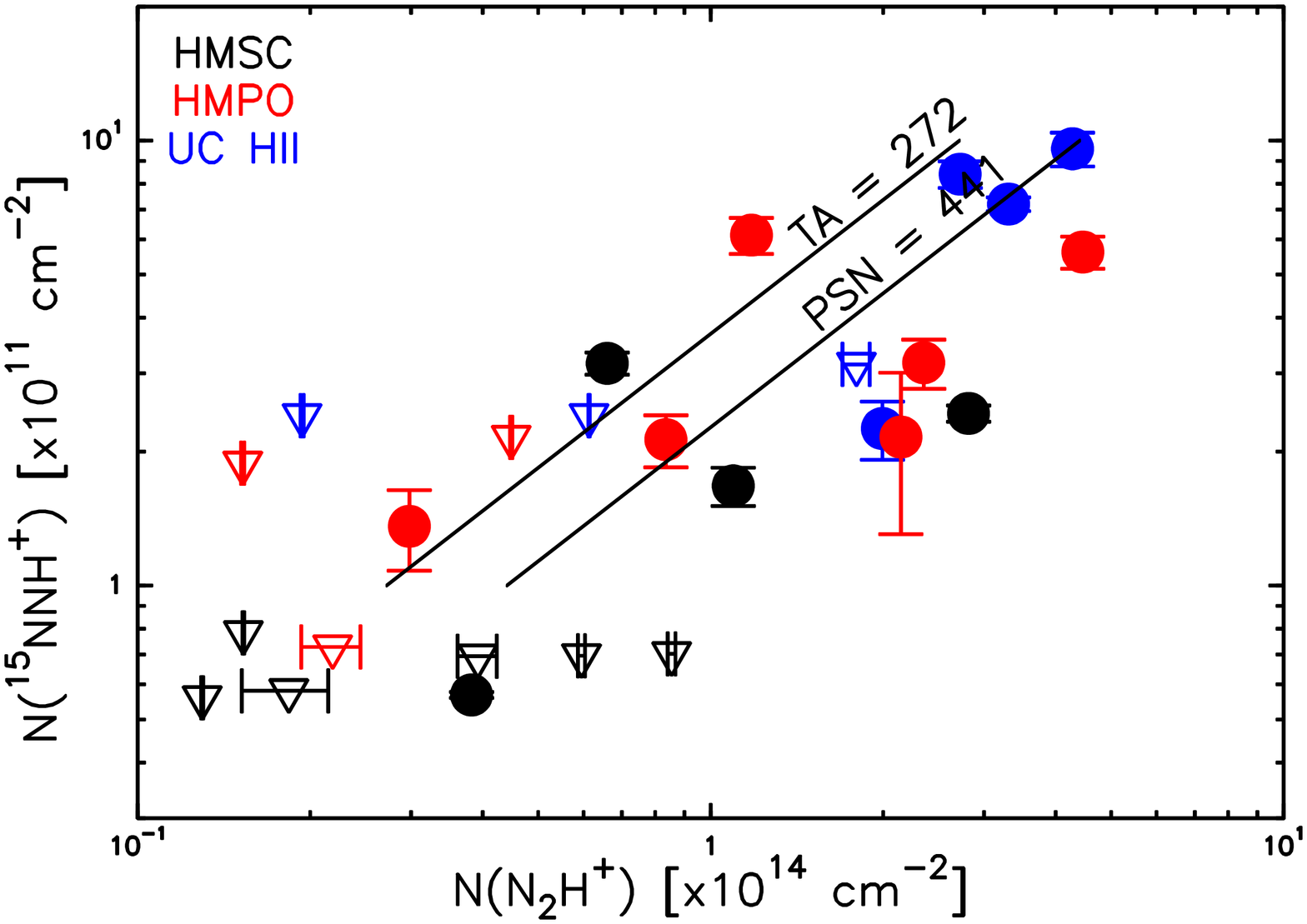}
                 \includegraphics[angle=-90,width=8cm]{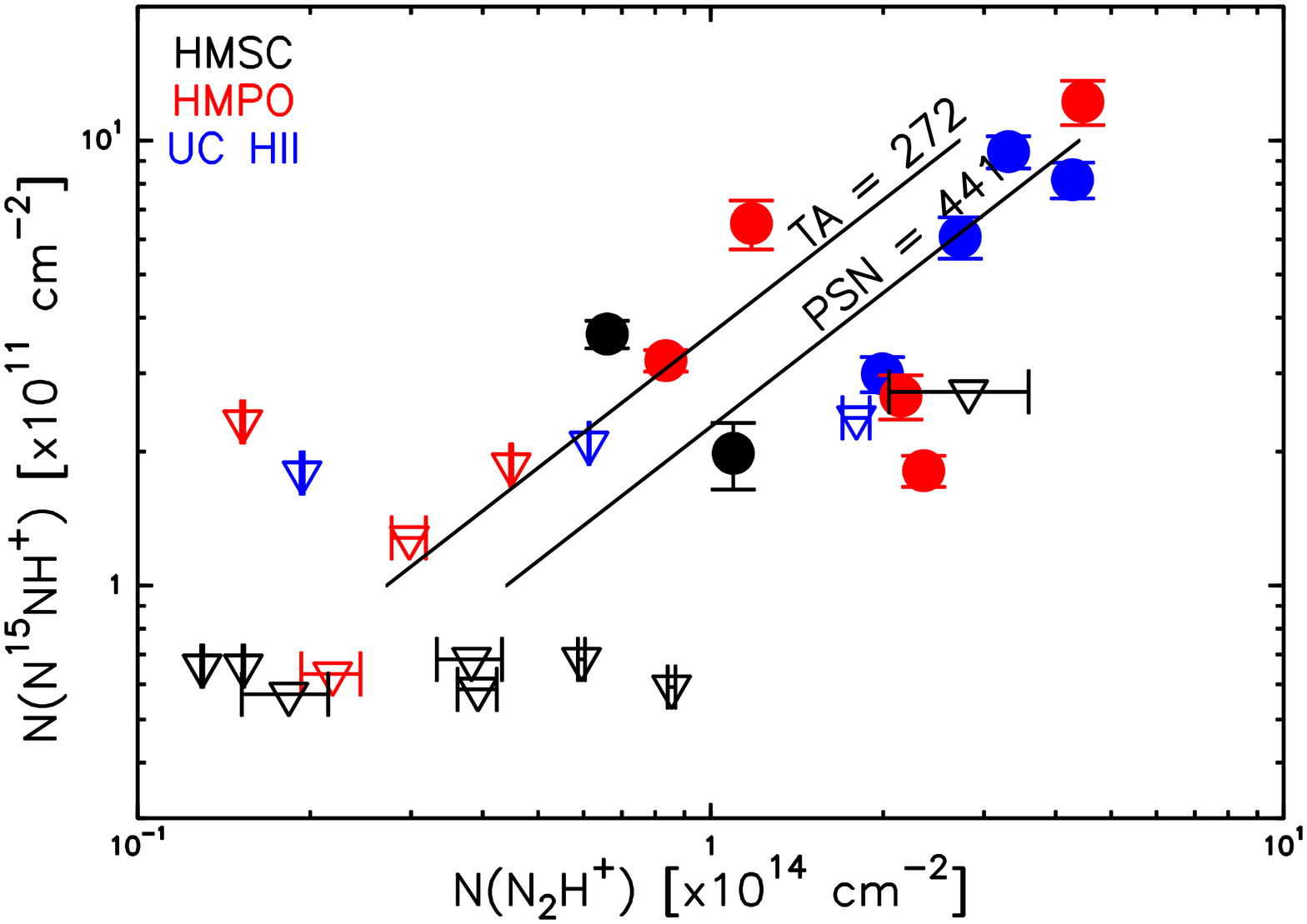}}
\centerline{\includegraphics[angle=-90,width=8cm]{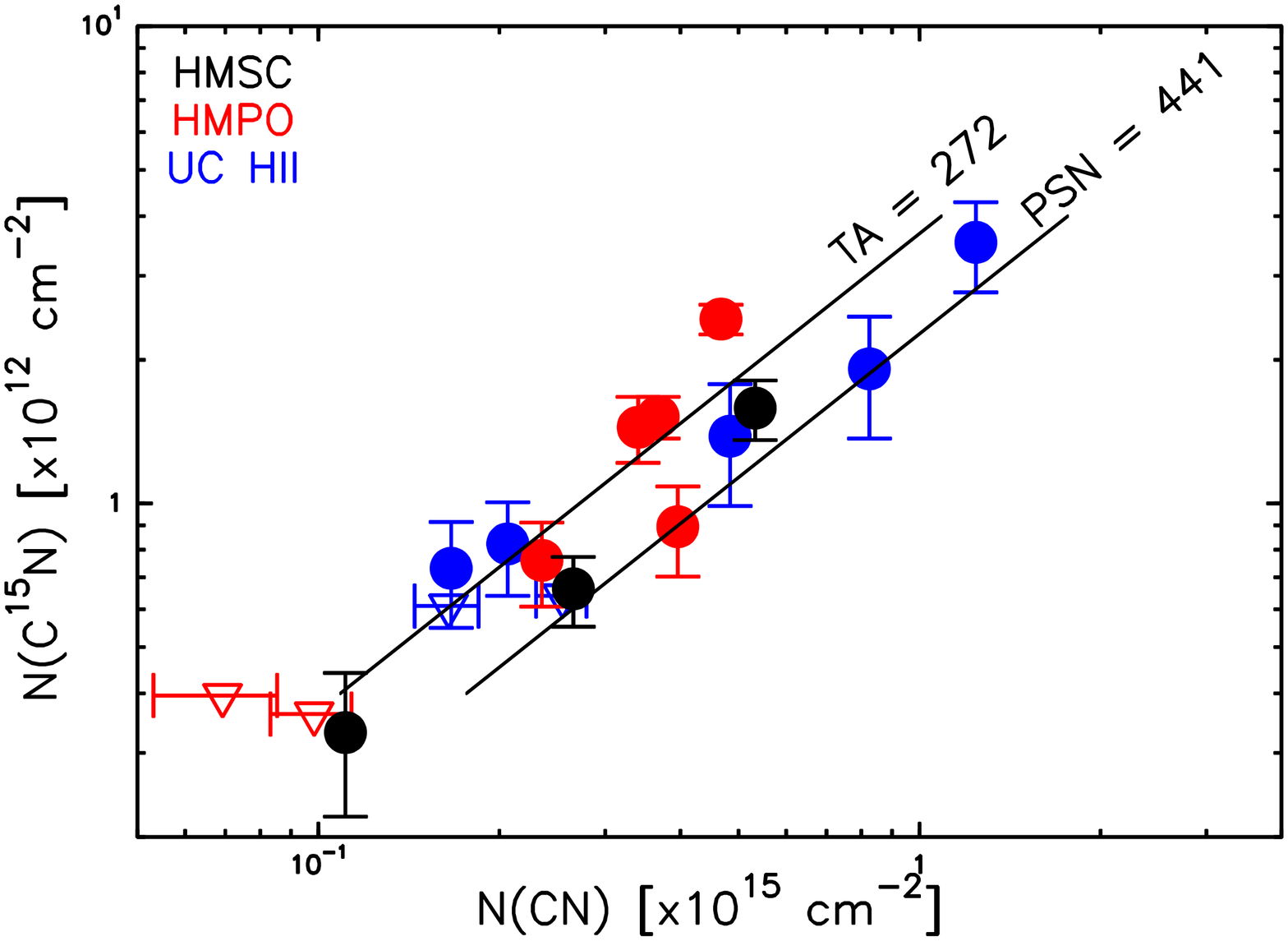}}
\centerline{\includegraphics[angle=-90,width=8cm]{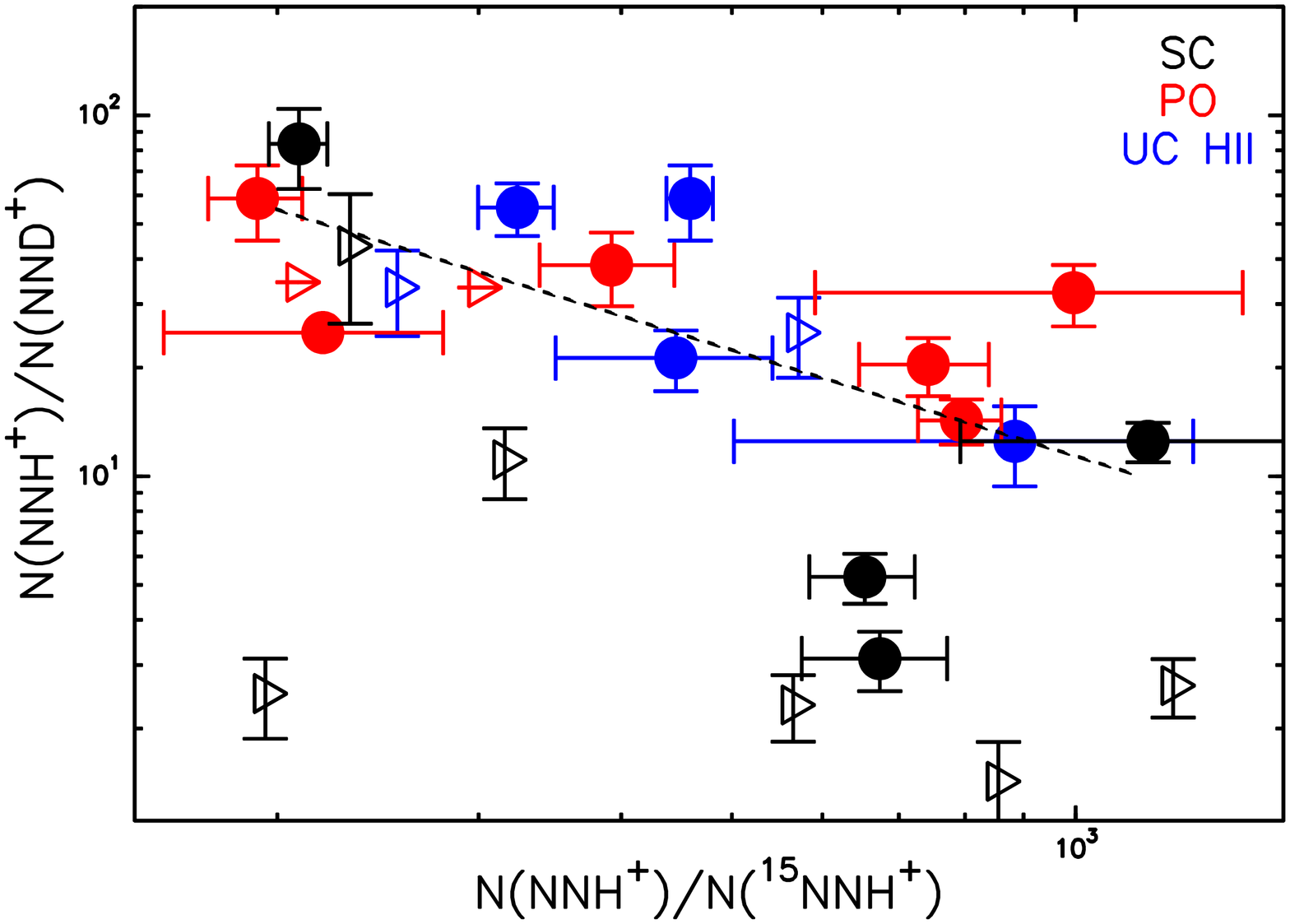}
                   \includegraphics[angle=-90,width=8cm]{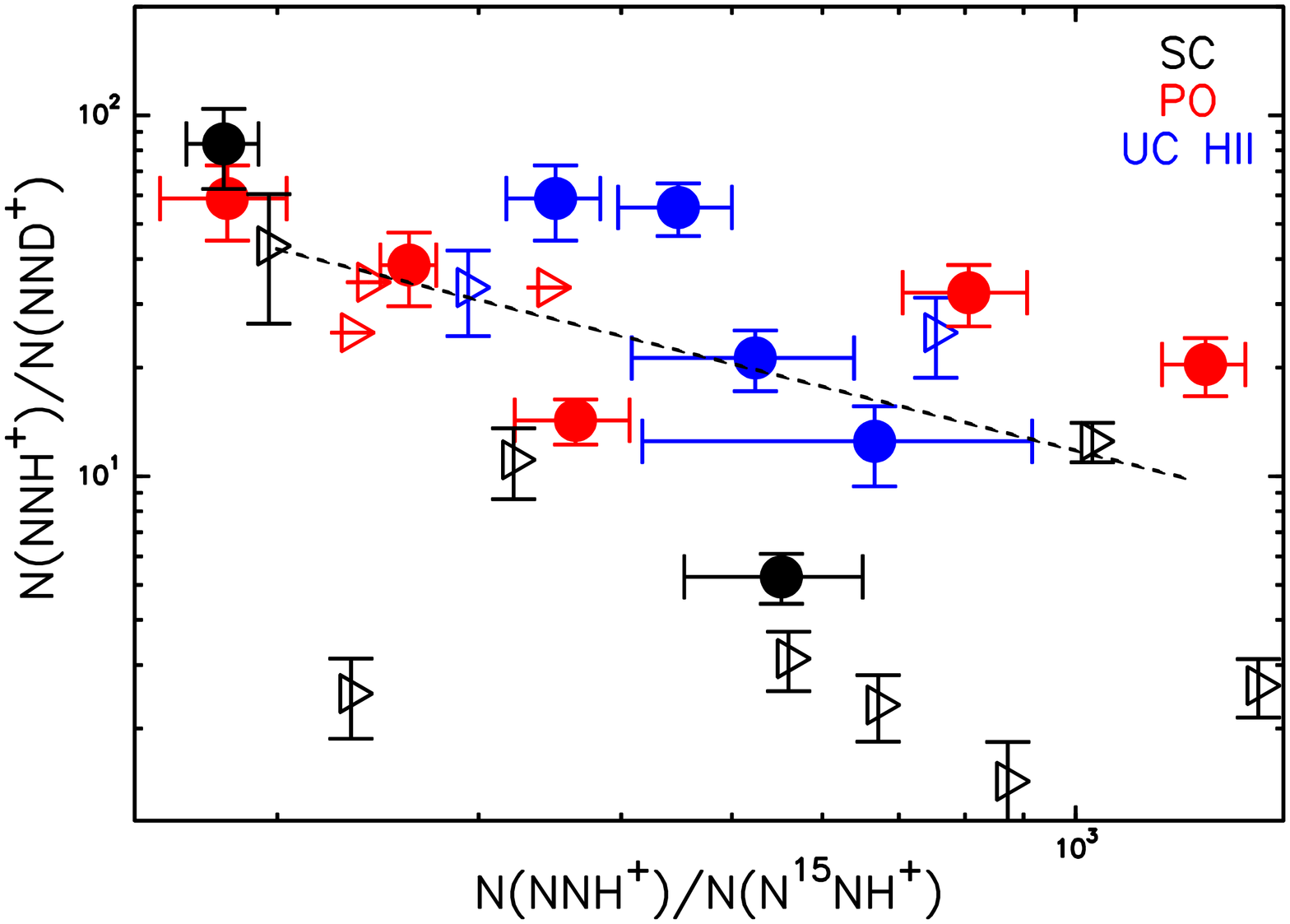}}
\caption[]{Top panels: column density of \H\ against that of \15N\ (left) 
and \N15\ (right). 
Middle panel: column density of CN against that of \CNII .
Bottom panels: comparison between the $^{14}$N/$^{15}$N isotopic ratio as derived 
from the column density ratios $N$(\H )/$N$(\15N ) (left) or 
$N$(\H )/$N$(\N15 ) (right), and the H/D isotopic ratio 
$N$(\H )/$N$(\D ). 
\newline
In all panels, the filled circles represent the detected sources (black = HMSCs; red = HMPOs;
blue = UC HIIs). The open triangles in the top and middle panels correspond to the upper 
limits on $N$(\15N ), $N$(\N15 ) or $N$(\CNII ), repectively, while in the bottom panels 
indicate lower limits on either $N$(\H )/$N$(\15N ) or $N$(\H )/$N$(\N15 ).
The solid lines indicate the mean atomic composition as measured
in the terrestrial atmosphere (TA) and in the protosolar nebula (PSN, see Hily-Blant et 
al.~\citeyear{hilyblant13} and references therein). The dashed lines in the bottom panels 
indicate least square fits.
}
\label{comp}
\end{figure*}

The comparison between the column densities of the $^{15}$N-containing 
species and those of their main isotopologues, derived as explained in 
Sects.~\ref{res_n2hp} and \ref{res_cn}, is shown in Fig.~\ref{comp}. 
The corresponding $^{14}$N/$^{15}$N isotopic ratios
are given in Tables~\ref{tab_res} and \ref{tab_res2}.
The first result emerging from these numbers and from Fig.~\ref{comp} is that the 
NNH$^+$/$^{15}$NNH$^+$ and NNH$^+$/N$^{15}$NH$^+$ 
ratios vary between 180 and 1300, namely over a range much larger than observed 
so far towards low mass objects in any N-bearing species. Some values are, therefore, 
compatible with both those of the TA and PSN, but much larger values are also 
measured. In addition, there is no evidence of a trend of these ratios with the 
evolution, because Fig.~\ref{comp} shows that the sources belonging
to different evolutionary stages possess similar $^{14}$N/$^{15}$N ratios, within
the uncertainties. Consistently, we do not find any correlation with the line widths,
which are thought to be good indicators of the evolutionary stage.
Therefore, whatever is the reason for this different $^{15}$N enrichment, 
time does not seem to play a role.

\begin{figure}
\centerline{\includegraphics[angle=-90,width=10cm]{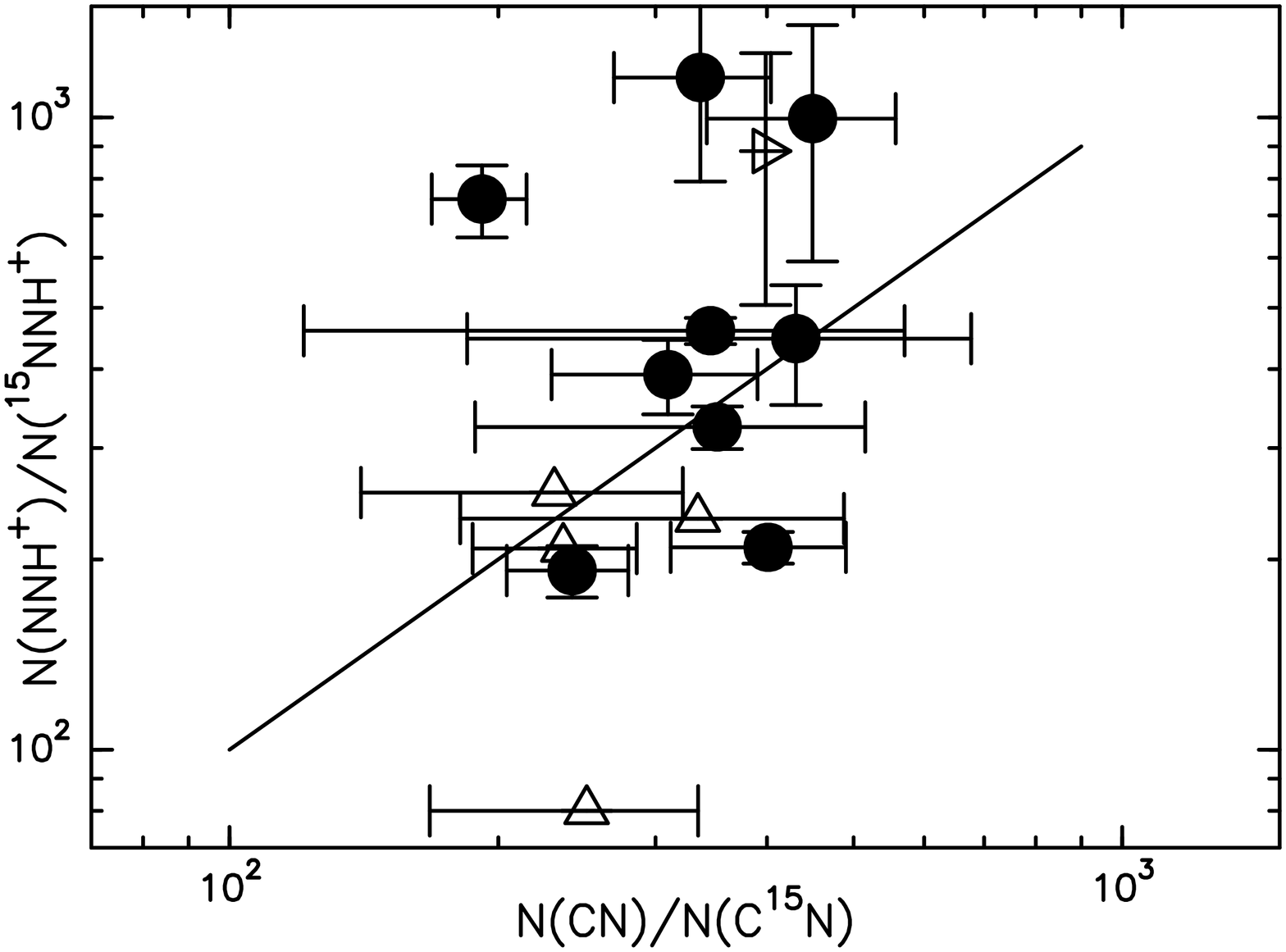}}
\caption[]{Comparison between the $^{14}$N/$^{15}$N isotopic ratio as derived 
from \H\ (y-axis) and CN (x-axis). On the y-axis we report only the ratio 
$N({\rm N_2H^+})/N({\rm ^{15}NNH^{+}})$, but the result is similar when using \N15.
The filled circles represent the cores detected in both \15N\ and \CNII. 
The open triangles with vertex pointing up correspond to lower limits 
on $^{14}$N/$^{15}$N from \H , the one with vertex pointing right
is the lower limit on $^{14}$N/$^{15}$N from CN for 23033+5951.
The straight line indicates the locus where $^{14}$N/$^{15}$N
from \H\ and CN are the same.
}
\label{nnhp_cn_comp}
\end{figure}

The second result is that the CN/C$^{15}$N ratio is within $\sim 190$ and $\sim 450$, 
namely consistent within the errors with the $^{15}$N enrichment of both TA and PSN. 
Interestingly, when the NNH$^+$/$^{15}$NNH$^+$ and CN/C$^{15}$N ratios 
can be measured in the same object, they have the same $^{14}$N/$^{15}$N ratio, 
within the error bars, only in nearly half of the sources detected in both \H\ and CN 
isotopologues, as shown in Fig.~\ref{nnhp_cn_comp}. 
However, globally the lower range of variation of the CN/C$^{15}$N could be a  
result biased by the fact that fewer sources were detected
in C$^{15}$N than in the \H\ isotopologues. Follow-up higher sensitivity 
observations will be helpful to address this aspect.

Note that, since the CN/C$^{15}$N has been derived by observations of $^{13}$CN, 
there is also the possibility that an error is introduced because of the variation of the 
$^{12}$C/$^{13}$C introduced by depletion of $^{13}$C (see Sect.~\ref{intro}).
However, the recent, very detailed study by Roueff et al. 
(\citeyear{roueff15}) predicts variations up to $\sim 20\%$
 (CN/$^{13}$CN ratio from 67 to 80), namely within the errors of our measurements. 
Thus, we conclude that our estimates of the CN/C$^{15}$N are substantially correct.

The measurements of both NNH$^+$/$^{15}$NNH$^{+}$ and CN/C$^{15}$N are at 
odds with the recent model by Roueff et al. (\citeyear{roueff15}), which predicts 
that the $^{14}$N/$^{15}$N should not vary in both species, unless the age of the 
condensation is shorter than 10$^5$ yrs. This possibility, however, seems to be ruled out by our 
observations, as the $^{14}$N/$^{15}$N is not different in the three evolutionary groups. 
We conclude that some important routes of nitrogen fractionation are still 
missing in the models. 

A hint of what they could be might be obtained considering the D-fractionation of 
NNH$^+$ towards the same sources where we measured the N-fractionation. 
The bottom panels in Fig.~\ref{comp} show NNH$^+$/$^{15}$NNH$^+$ and 
NNH$^+$/N$^{15}$NH$^+$ as a function of the NNH$^+$/NND$^+$, as previously 
measured by Fontani et al.~(\citeyear{fontani11}). Although not very strong, 
the data show a possible anti-correlation between the two parameters: 
the larger the $^{14}$N/$^{15}$N, the lower the H/D. 
Non-parametric statistical tests confirm this trend: the Kendall's $\tau $ rank
correlation coefficient is $\sim -0.5$ when the $^{14}$N/$^{15}$N is derived from
either \15N\ and \N15 , and 
the Pearson's correlation coefficient is $\sim -0.6$ and $-0.5$, respectively.
Moreover, the probability for a chance correlation is below the 2$\%$ in both cases.
Also, we have performed least square fits to the data, which provide 
slopes of $-0.98$ and $-0.80$ for the $^{14}$N/$^{15}$N ratios derived 
from \15N\ and \N15, respectively.
Based on this suggestive anti-correlation, we conclude that
the N-fractionation does not seem to be linked to the parameters that are 
known to affect the D-fractionation (if not to obtain the inverse effect): temperature and 
CO depletion (or density). 
In other words, the reactions leading to N-fractionation of 
NNH$^+$ and CN indeed do not have an energy barrier, as predicted by Roueff et 
al.~(\citeyear{roueff15}). 
Finally, and consistently with this last sentence, 
we do not find any correlation between the measured N-fractionation and 
the excitation temperature, contrary to the results of Wampfler et 
al.~(\citeyear{wampfler}), who claimed a tentative trend towards a 
temperature-dependent $^{14}$N/$^{15}$N fractionation.
However, because the anti-correlation found is not strong, care
needs to be taken in the interpretation of this result.
%

\section{Conclusions}
\label{conc}

We have observed the J = 1--0 rotational transitions of $^{15}$NNH$^+$, N$^{15}$NH$^+$ 
and \H , together with $N$ = 2--1 transitions of $^{13}$CN and C$^{15}$N, 
towards 26 massive star forming cores in different evolutionary stages, where the deuteration 
fraction has already been measured (Fontani et al.~\citeyear{fontani11}). 
We find $^{14}$N/$^{15}$N in CN between $\sim$ 190 and $\sim$ 450, and in 
N$_2$H$^+$ from $\sim$ 180 up to $\sim$ 1300. It is the first time that high 
$^{15}$N fractionations are found in N$_2$H$^+$.
However, in the fractionation process, time does not seem to play a role
because the $^{14}$N/$^{15}$N ratio does not depend on the evolutionary stage of the source.
We find a suggestive anticorrelation between D and $^{15}$N fractions, 
consistent with findings in Solar System material, where N isotope ratios largely
vary and only in a few cases they have been found to correlate with D/H ratios 
(Mandt et al.~\citeyear{mandt14}). This lack of correlation is also consistent with the
prediction of chemical models, which rule out any link between fractionation
of N and the causes of D enrichment, namely temperature and density (or CO depletion).

\section*{Acknowledgments}
FF and AP thank the IRAM staff for the precious help during the observations. 
FF is grateful to Fabien Daniel for the careful and constructive reading of the 
paper. PC acknowledges the financial support of the European Research Council
(ERC; project PALs 320620). AP acknowledges financial support from 
UNAM-DGAPA-PAPIIT IA102815 grant, M\'exico. This work has benefited from 
research funding from the European Community's Seventh Framework Programme.

{}

\end{document}